\documentclass[10pt]{article}

\usepackage{amsmath}
\usepackage{amssymb}

\usepackage{hyperref}

\usepackage{lineno}

\usepackage{microtype}
\DisableLigatures[f]{encoding = *, family = * }

\usepackage[dvips]{epsfig}
\usepackage[dvips]{graphicx}

\usepackage[labelfont=bf,labelsep=period,justification=raggedright]{caption}

\makeatletter
\renewcommand{\@biblabel}[1]{\quad#1.}
\makeatother
\usepackage[usenames]{color}

\date{}

\begin{document}

% Title must be 150 characters or less
\begin{flushleft}
{\Large
\textbf{Individual Biases, Cultural Evolution, and the Statistical Nature of Language Universals: The Case of Colour Naming Systems}
}
% Insert Author names, affiliations and corresponding author email.
\\
Andrea Baronchelli$^{1,\ast}$, 
Vittorio Loreto$^{2,3,4}$, 
Andrea Puglisi$^{5,2}$
\\
\bf{1} Department of Mathematics, City University London, London EC1V 0HB, UK
\\
\bf{2} Dipartimento di Fisica, Sapienza Universit\`a di Roma, P.le A. Moro 5, 00185 Roma, Italy
\\
\bf{3} Institute for Scientific Interchange (ISI), Via Alassio 11/c, 10126, Turin, Italy
\\
\bf{4} SONY Computer Science Lab (SONY-CSL), 5, Rue Amyot, 75005, Paris, France
\\
\bf{5} CNR-ISC, Piazzale Aldo Moro 5, 00185 Roma, Italy\\
$\ast$ E-mail: a.baronchelli.work@gmail.com
\end{flushleft}

% Please keep the abstract between 250 and 300 words
\section*{Abstract}
Language universals have long been attributed to an innate Universal Grammar. An alternative explanation states that linguistic universals emerged independently in every language in response to shared cognitive or perceptual biases. A computational model has recently shown how this could be the case, focusing on the paradigmatic example of the universal properties of colour naming patterns, and producing results in quantitative agreement with the experimental data. Here we investigate the role of an individual perceptual bias in the framework of the model. We study how, and to what extent, the structure of the bias influences the corresponding linguistic universal patterns. We show that the cultural history of a group of speakers introduces population-specific constraints that act against the pressure for uniformity arising from the individual bias, and we clarify the interplay between these two forces. 
%We believe that our simulations can help to shed light on the possible mechanisms at work in the evolution of language, and that they provide hints for further theoretical investigations. 
% Please keep the Author Summary between 150 and 200 words
% Use first person. PLOS ONE authors please skip this step. 
% Author Summary not valid for PLOS ONE submissions.   
%\section*{Author Summary}

\section*{Introduction}

\subsection*{Language universals and colour naming}
Different languages share a collection of structural properties, which are accordingly said to be \textit{universal} \cite{croft2002typology}. The origin and nature of this universality have been debated for decades in a dispute that is still far from being settled \cite{bickerton2009biological,smith2010evolution}. The classical view assumes that language learning requires innate language-specific knowledge \cite{chomsky1980rules,pinker1995language}. Every individual is endowed with a set of grammatical principles, defining a Universal Grammar, that are consequently found to be shared across all human languages. This view has however been questioned in various ways (see, for example, \cite{christiansen2008language}), and different computational approaches have shown that the hypothesis of a language-specific genetic endowment would result in a series of paradoxes \cite{chater2009restrictions}, while cultural transmission introduces informational bottlenecks that favour the emergence of regularities \cite{kirby2007innateness,griffiths2007language}. A systematic analysis has also shed new light on the very concept of universality \cite{evans2009myth}. Language universals must be intended in a statistical sense, rather than as necessary features of all human languages \cite{christiansen2008language,evans2009myth}. 

All of these more recent accounts, although heterogenous in other respects, agree that the observed cross-linguistic regularities emerged independently in every language in response to universal communicative, cognitive, or perceptual biases that are not specific to language \cite{deacon1998symbolic}. 
However, several issues remain open. Evans and Levinson  \cite{evans2009myth} have recently listed some of the more urgent questions, among which are: What are the biases responsible for a \textit{given} observed regularity? How do they generate a structure? What is the relation between the strength of a bias and the amount of cross-linguistic variability or universality? To this list we would add a question which is to us as urgent as the others: what is the origin of cross-linguistic variability \cite{baronchelli2012biological}? In other words, how effective is the specific history of a language in shaping specific, non universal properties?

Answering these questions is hard, but focusing on simple aspects of language has provided important insights. This is why colour categorisation has captured a great deal of attention over the last few decades. As observed in the World Color Survey (WCS) \cite{berlinkay,cook2005world}, different groups of individuals develop different colour naming patterns, but some universal properties can be identified by a statistical analysis over a large number of populations \cite{kay2003rqc}. Furthermore, it appears that basic colour names enter a language in a relatively fixed order \cite{berlinkay}. In general, the universality of colour naming is fully recognised as a genuine linguistic universal \cite{taylor2003lc,witkowski1977explanation,comrie1989language,deacon1998symbolic,gardner1985msn,lakoff-women,murphy2004bbc}, and has been the focus of many research efforts. In particular, several computational models have contributed to shedding light on the different theoretical hypotheses put forward in this context \cite{steels2005cpg,Belpaeme_Bleys2005,dowman2007ect,komarova2007emc,komarova2008pha,cg_pnas,jameson2009_1,jameson2009_2,cg_pnas_2,cg_pnas_vittorio,narens2012language,endrikhovski2002computational}. Even though most of these efforts have been devoted to modelling the emergence of shared colour categorisation patterns in a single population of individuals, more recent contributions have also addressed the issue of universality \cite{dowman2007ect,cg_pnas_2,cg_pnas_vittorio,xu2013cultural}.

%For example, Witowski and Brown have pointed out the importance of the findings in this domain for ``the lexical encoding of a wide range of phenomena in addition to colour"  \cite{witkowski1977explanation} (p.~56); Comrie has defended that ``Berlin and Kay's work also has more far-reaching implications for work on language universals and typology, and even for descriptive linguistics'', such as the one of ``providing one example of a psychological explanation of a linguistic universal" thanks to the identification of the role played by colour perception \cite{comrie1989language} (p.~38); Deacon has argued that the `cognitive bias' explanation of language universals naturally accounts for both the cross-linguistic properties exhibited by colour naming patterns, and the other linguistic universals, from morphology to syntax  \cite{deacon1998symbolic}.
%Color categorisation is therefore a prototypical example to understand language universals \cite{gardner1985msn,lakoff-women,murphy2004bbc}. With respect to most of the examples mentioned above, it has the advantage of being a well-defined problem, for which field data have been available for decades, and it continues to be the subject of intense investigation. 

\subsection*{The Category Game model}

Here, we focus on the Category Game model \cite{cg_pnas}, describing how a group of agents manages to establish a shared set of linguistic categories when faced with a continuous aspect of the environment, such as the hue channel of the colour spectrum.  According to the standard approach of language games \cite{wittgenstein53english,steels1995}, a population of agents can establish a shared category system through pairwise linguistic interactions alone. In every conversation one of the two agents tries to direct the attention of the other towards a specific object out of those that constitute the scene they are both facing. Depending on the success (or failure) of this operation both agents modify their internal state eventually leading to a global consensus. The model can be informed with a human psychophysiological parameter, namely the Just Noticeable Difference (JND), defining the minimum distance at which two stimuli from the same scene can be discriminated as a function of their wavelength \cite{bedford1958wavelength,long2006ssn}.  If the individuals are endowed with the human JND function, the statistical properties of the emerging categorisation pattern turn out to quantitatively match those observed in the World Color Survey data \cite{cg_pnas_2}, and the empirically observed colour hierarchy is reproduced \cite{cg_pnas_vittorio}. 

The agreement between the outcome of the model and experimental data corroborates the hypothesis according to which weak perceptual or cognitive biases can be responsible for the emergence of linguistic properties shared across independent languages \cite{deacon1998symbolic,christiansen2008language,evans2009myth}. However, in the model the bias does not constrain the structure of the shared categorisation pattern in a deterministic way. 
%Rather, its effect can be detected only through statistical tests run on a large number of languages. 
In fact, two forces shape the naming structure of a single population. Cultural history defines arbitrary consensus patterns that, once they have emerged, remain as frozen accidents and affect the further evolution of the language \cite{dunn2011evolved,mukherjee2011aging}, while the perceptual bias tends to prefer some patterns over others. At the level of a single population it is not possible to predict which force will prevail and to what extent. Statistically, however, it is possible to recover the signature of the bias.

Here, we investigate the role of the individual bias. We clarify, in the framework of the model, what a ``weak" bias is, how it works, and how it affects the cultural history of a linguistic pattern. To do so we exploit the flexibility offered by numerical simulations in two ways. First, we manipulate the nature of the bias, considering artificial JND functions, and comparing the impact they have on the emerging categorisation patters. We find that the human JND belongs to a class of special functions that allow for a good agreement with the experimental data, and that this agreement is by no means assured by any arbitrary JND. Then, we analyse in detail what ``weak" bias means by looking at the variability of the observed patterns. In particular, we study the cross-linguistic fluctuations of the categories which allow us to estimate the relative strength of cultural and cognitive or perceptual pressures in the framework of the model. We show that single languages can be driven far from universality by their own historical evolution, even though the vast majority of languages share, to some extent, universal properties. The peculiar nature of such fluctuations (characterised by {\em fat} tails, i.e., over-populated extreme events) suggests a relevant role for history-dependent correlations \cite{mukherjee2011aging}, which - in a sense - should be considered as a kind of {\em cultural pressure}. 
%Since, as mentioned above, the Category Game model is able to reproduce the statistical properties of the WCS data and the hierarchy of colour terms provided individuals agents are endowed with the human JND, we believe that our analysis may help to shed into the actual mechanisms responsible for the emergence of language universals.

\subsection*{Related work}

The role of weak biases has been investigated also in relation to their impact on cultural transmission \cite{boyd1988culture}. In particular, Kirby et al. \cite{kirby2007innateness} have investigated how a biological bias can translate into universal linguistic properties in the context of the Iterated Learning framework \cite{kirby2001spontaneous}, assuming that learners apply the principles of Bayesian inference \cite{mr1763essay}. The central result is that cultural transmission can magnify weak biases into strong universals, while strong biases could be shielded by transmission bottleneck, i.e., the limited amount of linguistic examples from which each speaker must learn the language. In the same framework, \cite{griffiths2007language} showed that linguistic structure can emerge when language learners use learning algorithms only slightly biased towards structured languages. As we mentioned above, and will see in the remainder of the paper, our work nicely complements these results by showing that the history of each language can introduce non-biological constraints as further biases shaping the language itself. While the Iterated Learning approach sheds light on the interplay between biology and cultural transmission in a formal way, we  tackle, from a different perspective, the relationship between biology and cultural evolution. 

Another interesting result concerns a numerical experiment by Regier et al \cite{regier2007color}. Here the hypothesis that colour naming represents optimal partitions of the colour space \cite{jameson199714} is tested through the definition of a measure of ``well-formedness" of the linguistic categories for colour. Optimal partitions of colour space are those that maximise this well-formedness, and the authors find that artificially generated colour-naming schemes that maximise this quantity in the full CIEL*a*b space resemble the colour naming scheme of some of the world's languages, even though ``many languages'' are not ``very similar'' to the hypothetically optimal model configurations. A second result is that given the colour naming scheme of any language, distortions away from that pattern often result in lower well-formedness. While this approach does not explain the observed number of categories, which is a parameter to be fixed before performing the analysis, it provides an interesting perspective on colour categorisation patterns. In particular, as the authors suggest, the fact that optimality does not account for all of the observed languages could be the result of cultural forces pushing languages far from the optimal partition. From this perspective, the Category Game helps to substantiate this argument by showing how culture and a universally shared parameter (here the candidate would be optimality) can interact. 

More recently, Xu et al. have shown that cultural transmission produces schemes of colour categorisation similar to those observed in the WCS in a laboratory experiment in which chain of individuals simulated the Iterated Learning scheme with a pre-determined number of terms \cite{xu2013cultural}. In particular, the authors borrow tools from information theory to show a significant match between their experiments and the WCS data. Also in this case, we believe that the Category Game represents a valuable complementary approach (i) offering a possible mechanism determining the cultural evolution of the small number of colour names observed in all the languages of the WCS, and (ii) suggesting with quantitative evidence that the underlying perceptual bias, magnified by cultural transmission in the Xu et al. experiment \cite{xu2013cultural}, could in fact be the human JND function. 

Finally, among the different computational models set forth to study colour categorisation \cite{steels2005cpg,Belpaeme_Bleys2005,dowman2007ect,komarova2007emc,komarova2008pha,cg_pnas,jameson2009_1,jameson2009_2,cg_pnas_2,cg_pnas_vittorio,narens2012language,endrikhovski2002computational}, it is important to mention the one introduced in \cite{dowman2007ect}, which replicates the frequency of colour terms observed in the WCS by adopting an evolutionary perspective, along with a Bayesian scheme of term acquisition, and dealing with a one-dimensional colour space relative to hue dimension (as the Category Game does). In \cite{dowman2007ect} the hue space is discrete, featuring $40$ possible hue values, and the model assumes (i) that the universal foci are predefined and unevenly spaced in the colour space and (ii) that  colour terms denote a contiguous range of colours. In the Category Game, on the contrary, the hue space is continuous and the two hypotheses above are not necessary. In particular, the existence of universal foci is a byproduct of the fact that the individuals are endowed with the human JND function, while the fact that colour names refer to contiguous frequencies is an emerging property of the dynamics \cite{cg_pnas}.

Considered together, the two models of \cite{dowman2007ect} and \cite{cg_pnas} indicate that at least some properties of the WCS data can be accounted for by considering the hue channel alone. Specifically, this is true for basic colour terms, while it might not be the case for the naming of composite or derived colours \cite{kay1999asymmetries}. Interestingly, the crucial role played by hue has recently found a biological rationalisation in the neurological study of primate vision. According to Xiao et al., the fact that the non-random properties of colour-naming patterns can be accounted for by the wavelength JND, as shown by \cite{cg_pnas_2}, implies a link between the universal constraints and the functional characteristics of the retinogeniculate pathway \cite{xiao2011biological}. And in fact, the study carried on by the same authors establishes a direct link between a universal constraint on colour naming, namely the fault line close to the ``warm"-``cold" colour distinction, and the cone-specific information that is represented in the primate early visual system, specifically the L- versus M-cone contrast \cite{xiao2011biological}. 

The Category Game has in fact been extended to more dimensions in order to describe the evolution from a mostly brightness-based to a mostly hue-based colour term system comparable to what happened during the Middle English period in response to the rise of dyeing and textile manufacturing \cite{cg_2d}. The 2D model is obviously more complex and requires further specifications of the rules determining the individual categorisation process. The outcome is richer than its one-dimensional counterpart, but the unavoidable consequence is that it is less transparent to interpretation. Since our aim here is to clarify the issue of the interplay between a cognitive or perceptual bias and the universality properties of the generated languages, and that the latter happens to appear already in the 1D model, we stick to the simplest case (see \cite{frankfurt} for a deeper discussion on why choosing simplicity when confronted with such a choice).

% You may title this section "Methods" or "Models". 
% "Models" is not a valid title for PLoS ONE authors. However, PLoS ONE
% authors may use "Analysis" 
\section*{Materials and Methods}

The essential features of the Category Game model are:

\begin{enumerate}
\item It considers the categorisation of a continuum perceptual channel \cite{cg_pnas}; 
\item It generates categorisation patterns whose statistical properties are in quantitative agreement with the ones observed in the WCS data \cite{cg_pnas_2};  
\item It reproduces the hierarchy of colour terms \cite{cg_pnas_vittorio}. 
\end{enumerate}

\noindent In this section we briefly recall the model definition and some previous results. 
\subsection*{The model}
The Category Game involves a population of $N$ artificial agents. Starting from scratch and without any pre-defined colour categories, the model dynamically leads, through a sequence of pair-wise interactions (games), to the emergence of a highly shared set of linguistic categories of the visible light spectrum. The parameters of the model are the population size $N$ and the JND function. 

For the sake of simplicity and without loss of generality (see also \cite{cg_2d}), colour perception is reduced to a single analogical continuous perceptual channel, each light stimulus being a real number in the interval $[0,1)$, which represents a rescaled wavelength. A categorisation pattern is identified with a partition of the interval $[0,1)$ in sub-intervals, or perceptual categories. Individuals have dynamical inventories of form-meaning associations linking perceptual categories with their linguistic counterparts, basic colour terms, and these inventories evolve through elementary language games. At each time step, two players (a speaker and a hearer) are randomly selected from the population and a scene of $M > 1$ stimuli is presented. Two stimuli cannot appear at a distance smaller than JND$(x)$ where $x$ is the value of one of the two. This is the way in which the JND is implemented in the model. On the basis of the presented stimuli, the speaker discriminates the scene, refining if necessary its perceptual categorisation, and utters the colour term associated with one of the stimuli. The hearer tries to guess the named stimulus, and based on their success or failure, both individuals rearrange their form-meaning inventories. New colour terms are invented every time a new category is created for the purpose of discrimination, and are spread through the population in successive games. A detailed description of the model is presented in the Appendix.

The dynamics proceeds as follows \cite{cg_pnas}. At the beginning all individuals have only the perceptual category $[0,1)$ with no associated name. During the first phase of the evolution, the pressure for discrimination makes the number of perceptual categories increase, resulting in widespread synonymy due to the many different words used by different agents for similar categories. This kind of synonymy reaches a peak and then dies out, in a similar way as in the Naming Game \cite{steels1995,baronchelli_ng_first}. When on average only one word is shared by the whole population for each perceptual category, a second phase of the evolution starts. During this phase, words expand their dominion across adjacent perceptual categories, merging several perceptual categories giving rise to a new type of categories, namely the ``linguistic categories".  The structure of the linguistic categories evolves through a domain growth process, known in statistical physics as coarsening \cite{bray94}.  Their number is progressively reduced till the system undergoes a dynamical arrest, featuring a slowing down of the domain growth, along much the same lines as the physical processes by which supercooled liquids approach the glass transition \cite{cavagna2009supercooled}. 

In this long-living, almost stable, phase, usually after $10^4$ games per player, the linguistic categorisation patterns of the individuals overlap with each other to an extent ranging between $90\%$ and $100\%$. The success rate and the similarity of category patterns across different agents remain stable for a time of $\sim 10^6$ games per player, and this pattern is considered as the final categorisation pattern generated by the model, to be compared with human colour categories (see below). 
% If one waits for a much longer time (diverging with $N$), the number of linguistic categories is observed to drop down: this non-realistic effect is due to the slow diffusion of category boundaries.
Note that, at the level of the Category Game, categories can be equivalently described in terms of boundaries or prototypes. 
%Slow diffusion of boundaries ultimately takes place due to small size effects. 
%Recent investigations have demonstrated that this phase can occur on very long time-scale, with autocorrelation properties typical of an aging material, such as a glass. 
The shared pattern in the long stable phase between $10^4$ and $10^6$ games per player is the main subject of the experiment described in the following section. As already observed in \cite{cg_pnas}  the number of linguistic colour categories observed in this phase is of the order of $15 \pm 10$, as in natural languages, even though discrimination allows for the existence of hundreds of categories.

\subsection*{Comparison with real-world data}

A large amount of data on colour categorisation was gathered in the World Color Survey \cite{berlinkay,kay2003rqc}, in which individuals belonging to different cultures had to name a set of colours. 
%The main finding is that colour systems across language, far from being random, exhibit instead certain statistical regularities. 
In this section, we review how the Category Game model has been used to reproduce some of the most important features of the empirical data. 
%run a \textit{Numerical} World Color Survey that produced results in quantitative agreement with the experimental ones \cite{cg_pnas_2}.

The fundamental source of experimental data on colour naming systems comes from the World Colour Survey. P. Kay and B. Berlin \cite{berlinkay} ran a first survey on 20 languages in $1969$. From 1976 to 1980, the enlarged World Color Survey was conducted by the same researchers along with W. Merrifield \cite{kay1991biocultural} and the data have been publicly available since 2003 on the website {\tt http://www.icsi.berkeley.edu/wcs}. These data concern the basic colour categories in $110$ languages without written forms and spoken in small-scale, non-industrialised societies. On average, $24$ native speakers of each language were interviewed. Each informant had to name each of $330$ colour chips produced by the Munsell Color Company that represent $40$ gradations of hue and maximal saturation, plus $10$ neutral colour chips (black-gray-white) at $10$ levels of value. The chips were presented in a predefined, fixed random order, to the informant who had to tag each of them with a ``basic colour term" in her language (for more details see \cite{berlinkay}).

After two decades of intense debate \cite{gardner1985msn}, Kay and Regier \cite{kay2003rqc} performed a quantitative analysis proving that the colour naming systems obtained in different cultures and language are in fact not random. Through a suitable transformation, they identified the most representative chip for each colour name in each language and projected it into a suitable metric colour space (namely, the CIEL*a*b* colour space). To investigate whether these points are more clustered across languages than would be expected by chance, they defined a dispersion measure on this set of languages $S_0$

\begin{equation}
D_{S_0}= \sum_{l,l^* \in S_0} \sum_{c \in l} \mbox{min}_{c^* \in
  l^*} \mbox{distance}(c,c^*),
  \label{eq:disp}
 \end{equation}
 
 \noindent where $l$ and $l^*$ are two different languages, $c$ and $c^*$ are two basic colour terms respectively from these two languages, and $\mbox{distance}(c,c^*)$ is the distance between the points in colour space in which the colours are represented. To give a meaning to the measured dispersion $D_{S_0}$, Kay and Regier created ``new" datasets $S_i$ ($i=1,2,..,1000$) obtained through random rotations of the original set $S_0$, and measured the dispersion of each new set $D_{S_i}$. The human dispersion appears to be distinct from the histogram of the ``random'' dispersions with a probability larger than $99.9\%$. As shown in Fig. 3A of \cite{kay2003rqc}, the average dispersion of the random datasets, $D_{neutral}$, is $1.14$ times larger than the dispersion of human languages. Thus, human languages are more clustered, i.e., less dispersed, than their random counterparts, confirming the existence of some kind of universality.
 
Another important finding concerning the World Colour Survey is the existence of a hierarchy of basic colour names that were adopted  by individual cultures in a relatively fixed order  \cite{berlinkay}. Thus, basic colour terms can be organised into a hierarchy around the focal colours black, white (where black and white usually map meaning close to the general panchromatic English terms dark and light or dull and brilliant rather than equivalent to the specific achromatic terms black and white), red, green, yellow, and blue always appearing in this order across cultures in such a way that if a culture has, for example a word for red, it also has a name for black and white (but not vice-versa), if it has a name for green it also has a name for red (but not vice-versa) etc.    

\subsubsection*{The Numerical World Color Survey}

\begin{figure} [!h]
\begin{center} \includegraphics*[width=0.55\textwidth]{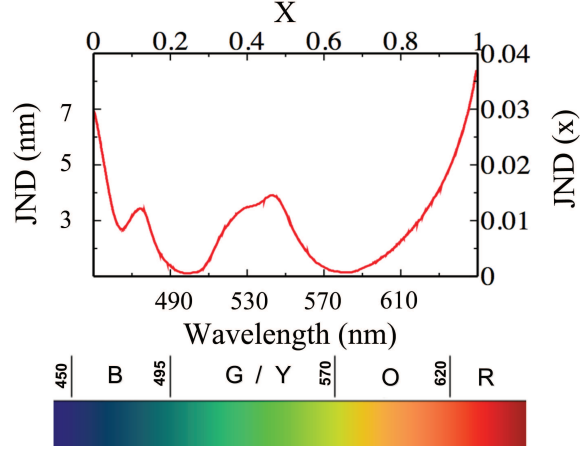} \end{center} 
\caption{{\bf The human Just Noticeable Difference (JND) function}  describes the wavelength change in a monochromatic stimulus needed to elicit a particular JND in the hue space. Both as a function of the wavelength of the incident light (measured in nanometers) and on the rescaled interval $[0,1)$. For convenience we display also the spectrum of the visible light. For the purpose of the Category Game we rescale the monochromatic stimulus in visible spectrum (measured in nanometers) in the range $[0,1)$ for the Topic $x$. In the same way the JND function (in nanometers in the left y-axis) is rescaled into a $JND(x)$ function (right y-axis).}
 \label{f:human_jnd}
 \end{figure}

The core of the analysis described above is the comparison of the clustering properties of a set of \textit{true} human languages against the ones exhibited by a certain number of randomised sets. In replicating the experiment it is therefore necessary to obtain two sets of synthetic data, one of which must have some human ingredient in its generation. The idea put forth in \cite{cg_pnas_2} is to act on the JND function. Human beings are endowed with a JND for the hue that is a function of the wavelength of the incident light (see Fig.~\ref{f:human_jnd}). This is the only parameter of the model encoding the finite resolution power of perception, or equivalently the human Just Noticeable Difference.

Starting from the human JND, different artificial sets can be created:

\begin{itemize}
\item \textit{``Human"} categorisation patterns are obtained from populations whose individuals are endowed with the rescaled human JND;
\item \textit{Neutral} categorisation patterns are obtained from populations in which the individuals have constant JND (JND$=0.0143$), which is the average value of the human JND (as it is projected on the $[0,1)$ interval).
\end{itemize}

\noindent In analogy to the WCS experiment, the randomness hypothesis in the NWCS for the neutral test-cases is supported by symmetry arguments: in neutral simulations there is no breakdown of translational symmetry in the colour space, which is the main bias in the``human" simulations.

Thus, the difference between ``human" and neutral data originates from the perceptual architecture of the individuals of the corresponding populations. A collection of ``human" individuals form a ``human" population, and will produce a corresponding ``human" categorisation pattern. In a hierarchical fashion, finally, a collection of populations is called a \textit{world}, which in \cite{cg_pnas_2} is formed either by all ``human" or by all non-``human" populations. For each world, the value of the dispersion $D$, defined in Eq.~(\ref{eq:disp}), is calculated, in order to quantify the amount of dispersion of the languages (or categorisation patterns) belonging to it. In the actual WCS there is of course only one human World (i.e., the collection of $110$ experimental languages), while in \cite{cg_pnas_2} several (i.e., $1500$) worlds were generated to gather statistics both for the ``human" and non-``human" cases.

The main results of the NWCS is that the Category Game, informed with the human $JDN(x)$ curve, produces a class of ``worlds'' featuring a dispersion lower than and well distinct from that of the class of ``worlds'' endowed with a non-human, i.e., uniform, $JND(x)$. Moreover, the ratio observed in the NWCS between the average dispersion of the ``neutral worlds'' and the average dispersion of the ``human worlds'' is $D_{neutral} / D_{human} \sim 1.14$, very similar to the one observed between the randomised datasets and the original experimental dataset in the WCS. Crucially, these findings are robust against changes in such parameters as the population size $N$, the distribution of the stimuli, the number of objects in a scene $M$, and the time of measurement (as long as measures are taken in the temporal region in which a categorisation pattern exists) \cite{cg_pnas_2}.

It is worth noting that often the outcome of a multi-agent model is employed to test the viability of theoretical hypotheses at a more abstract level \cite{frankfurt}. In this case, the results of the NWCS compare with WCS data not only from a qualitative point of view, but also quantitatively. Moreover, the very design of the model suggests a possible mechanisms lying at the roots of the observed universality, previously formulated on the basis of theoretical analysis (see, for instance \cite{deacon1998symbolic,christiansen2008language}). Human beings share certain perceptual biases that, even though not so strong as to deterministically affect the outcome of the categorisation process, are capable of influencing category patterns. However, this influence can be made evident only through a statistical analysis performed over a large number of languages, since any two (or few) languages could provide a misleading signal, either in the direction of strong universality or on the contrary of a total lack of shared features. 

\subsubsection*{Reproducing the hierarchy of colour names}

To address the problem of the origin of a hierarchy of colours the Category Game has been generalised to allow for the emergence of a series of shared linguistic layers each of which could guarantee communicative success in progressively more complex tasks \cite{cg_pnas_vittorio}. In the generalised version a higher and more refined linguistic layer is accessed by the agents only if in a game both the topic and the object have the same name and thus, there is a `failure' to differentiate between the two, possibly resulting in a failure in communication, referred as ``failure with name''. While the first layer of linguistic categories (Level 0, the one present also in the original Category Game) can be likened to the emergence of primary colour names, the successive layers (Levels 1,2,...) might be linked to the emergence of complex colour names when the knowledge of the primary colour names is not enough to achieve a reasonable communicative success (one can think of a linguistic community comprising specialised individuals, for instance painters, textile and cosmetic manufacturers \cite{plumacher_2007}). Even though a detailed discussion of this generalisation of the Category Game is out of the scope of this paper (we refer for this to \cite{cg_pnas_vittorio}), it is interesting to summarise the main results. 

A first observation concerns the frequency of access to higher levels of linguistic categorisation as a function of the local value of the JND. It turns out that the agents need to access the higher level early in regions (in the hue space) corresponding to high values of the JND, while they access it quite late in regions corresponding to low values of the JND. This indicates that an agreement at Level 0 is reached faster in regions with high values of the JND resulting in more cases of  ``failure with name'' in these regions, thereby, forcing the agents to access Level 1. Starting from this observation it is interesting to compute the time needed to get consensus in different regions of the hue space corresponding to different values of the JND. It turns out that the emergence of consensus occurs first in regions corresponding to high values of the JND while it occurs last in regions corresponding to very low JND. Strikingly, if the regions are arranged according to the time to reach a desired level of consensus, then they get organised into a hierarchy with [red, (magenta)-red], [violet], [green/yellow], [blue], [orange] and [cyan] (or [cyan] and [orange] as is usually observed for secondary basic colour names) appearing in this order. The names on which faster agreement is reached turn out to be the basic colour names that first emerge in a population, in excellent quantitative agreement with the empirical observations of the World Colour Survey. While we will not use the generalised model in this paper, this finding confirms the validity of the Category Game framework and and its ability to compare to and reproduce the empirical data of the WCS.   

% Results and Discussion can be combined.
\section*{Results}  
\label{sec:original}

In this section we investigate the role of the perceptual bias. First, we check to what extent the very good agreement between the outcome of the model and the WCS data can be related to the specific structure of the human JND. We show that different instantiations of the JND bias significantly alter the agreement between simulations and true data. 
%We point out that a single feature of the JND function, namely its roughness, controls the nature of the emerging universality\footnote{Generally, one refers to the term roughness to quantify the vertical deviations of a generic function (in our case the JND function) with respect to an ideal curve (in our case its constant average value). Here, we do not enter in further details and avoid to make the argument quantitative because this would go beyond the aim of this paper. When we refer to the term roughness we shall mean a qualitative notion of roughness.}. 
This result rules out the possible criticism that any kind of bias would lead to the same kind of universality in the model. 

Then, we compare the categorisations of different simulated populations, and we measure the dispersion of their colour naming patterns in order to study the interplay between the bias and the stochasticity introduced by the underlying cultural process. We show that the stochastic cultural process can generate shared category patterns that affect the ensuing evolution, effectively competing with the cross-cultural ordering tendency of the bias.  This explain why different populations can present significantly different categorisation patterns even when all individuals share the same JND function. We argue that our results allow us to better understand what a ``weak" bias is, and to what extent it may affect the emergence of the universal properties of language.

\subsection*{Artificial JNDs}

\begin{figure}
\begin{center} \includegraphics*[width=0.55\textwidth]{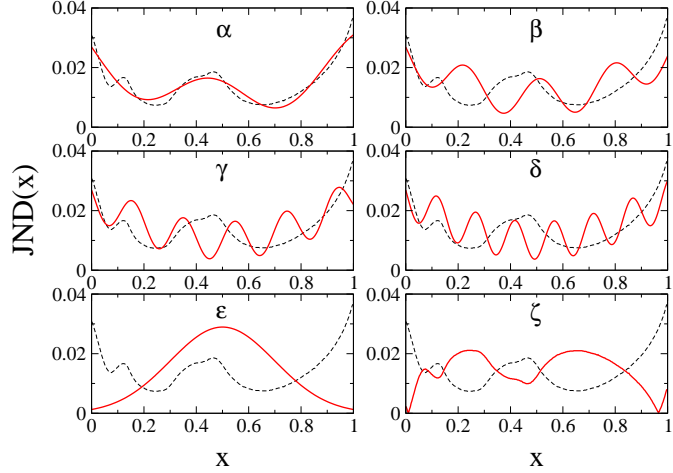} \end{center} 
\caption{{\bf Artificial JNDs.} Each panel depicts one of the artificial JNDs used in the experiments (continuous lines). The empirical fit of the human JND (dashed line, all panels) is obtained according to the expression JND$(x) = c_1 + c_2 \cos(ax+c_3) + c_4 (x-0.5)^2$, where $c_1,c_2,c_3,c_4$ and $a$ are fitting constants (case $\alpha$). By constant increases of the $a$ parameter one finds increasingly irregular artificial JNDs ($\beta, \gamma$ and $\delta$). The last two panels present a Gauss-like JND ($\epsilon$), and a reflection of the JND around the average value (with a further prescription avoiding negative values of the JND) ($\zeta$).}
 \label{f:jnds}
 \end{figure}

%\begin{figure}[t]
%\begin{center}
%\includegraphics*[width=0.55\textwidth]{artificial_jnds.eps}
%\end{center}
%\caption{{\bf Artificial JNDs.} Each panel depicts one of the artificial JNDs used in the experiments (continuous lines). Case $\alpha$ is the empirical fit of the human JND (dashed line, all panels) obtained according to the expression JND$(x) = c_1 + c_2 \cos(ax+c_3) + c_4 (x-0.5)^2$, where $c_1,c_2,c_3,c_4$ and $a$ are fitting constants. Cases $\beta, \gamma$ and $\delta$ are obtained by constant increases of the $a$ parameter. Curve $\epsilon$ is a Gaussian-like JND, while case $\zeta$ is a reflection of the JND around the average value (with a further prescription avoiding negative values of the JND). }
% \label{f:jnds}
% \end{figure}
% 
Only two JND functions have been tested in previous works, namely the human and the flat uniform JNDs, in order to contrast the naturalistic case with an unbiased neutral one. Here, we exploit the opportunity offered by numerical simulations to test different, artificial, JNDs. The aim of this new experiment is to understand whether the agreement with data is related to the specific shape of the human JND or rather if it can be achieved, in the framework of the model, with any JND. Of course, any bijettive function defined in $[0,1)$ could play the role of a JND, and a systematic exploration is therefore unfeasible, nor would it be useful. Rather, we start from the human JND, fit it empirically, and then modify the fitted function so as to progressively diverge from the original one. In particular, we insert local extreme points (minima and maxima) to increase the roughness of the JND. Furthermore, we consider also a Gauss-shaped JND as well as an inverted human JND. Fig.~\ref{f:jnds} sketches the adopted functions. The key quantity studied here is, again, the dispersion $D$ defined in Eq.~(\ref{eq:disp}) among languages, which is the natural measurement (complementary) of the degree of similarity, as already discussed in the literature~\cite{kay2003rqc}.

\begin{figure}
\begin{center} \includegraphics*[width=0.55\textwidth]{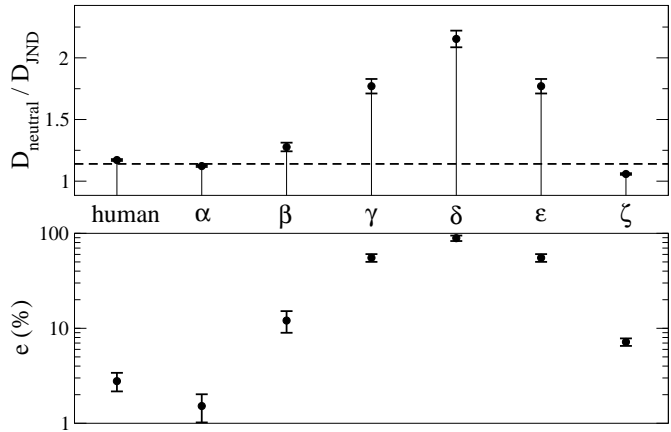} \end{center} 
\caption{{\bf Numerical World Color Survey with artificial JNDs.}
  Top: the dispersion of the JND case, as defined in Eq. \ref{eq:disp}, normalised by the neutral one obtained with a flat JND is
  plotted for populations of size $N=100$. The horizontal line indicates the
  experimental value $1.14$ obtained by the analysis of WCS data in
  \cite{kay2003rqc}.  Bottom: the relative error $e(\%)$
  between the average dispersions and the experimental result,
  $e=|sim-exp|/exp$ is plotted (in $\%$). This measure quantifies the distance of the dots from the horizontal line in the Top panel.  Different JNDs are named
  after Fig.~\ref{f:jnds}. Vertical bars refer to the variation of values
  in the late stage of the simulation, in the range $1.5\times 10^6-2
  \times 10^6$ games per agent.  }
 \label{f:nwcs_jnds}
 \end{figure}

Fig.~\ref{f:nwcs_jnds} shows the outcome of the numerical experiments run with several artificial JNDs. It is clear that the human JND, along with slight variations of it, performs well, while more irregular functions (obtained according to the expression JND$(x) = c_1 + c_2 \cos(ax+c_3) + c_4 (x-0.5)^2$, where $c_1,c_2,c_3,c_4$ and $a$ are fitting constants) weaken the agreement with the empirical data (Cases $\beta, \gamma$ and $\delta$ are obtained by constant increases of the $a$ parameter). The Gaussian-like function (case $\epsilon$) and the inverse JND (case $\zeta$), perform worse than the human JND even thought not dramatically badly. It is worth stressing that the relative error between the results obtained with the simulated JND and the experimental one of \cite{kay2003rqc} ($e=|sim-exp|/exp$) is plotted in logarithmic scale (Fig.~\ref{f:nwcs_jnds}, bottom) and that an error below $e=10\%$ can very well be considered as noticeably small given the simplicity of the Category Game model.

%\begin{figure}[t]
%\begin{center}
%%\includegraphics*[width=0.55\textwidth]{fig_ratios_jnd2_PUGLIO.eps}
%\end{center}
%\caption{{\bf Numerical World Color Survey with artificial JNDs.}
%  Top: the dispersion of the JND case normalized by the neutral one is
%  plotted for populations of size $N=100$. Bottom: the relative error
%  between the average dispersions and the experimental result,
%  $e=|sim-exp|/exp$ is plotted (in $\%$). Different JNDs are named
%  after Figure~\ref{f:jnds}. The horizontal line indicates the
%  experimental value $1.14$ obtained by the analysis of WCS data in
%  \cite{kay2003rqc}. Vertical bars refer to the variation of values
%  in the late stage of the simulation, in the range $1.5\times 10^6-2
%  \times 10^6$ games per agent.  }
% \label{f:nwcs_jnds}
% \end{figure}
 
Overall, smooth JNDs produce a weaker clustering of the colour naming patterns across different populations, and vice versa rough JNDs force a greater uniformity. 
%This allows us to conclude that roughness is a parameter determining the strength of the bias. 
The amount of universality observed in a set of languages depends therefore (also) on a specific property of the individual bias. Speculatively, this result suggests that, had the human JND been less smooth, we would have observed a greater regularity in colour naming patterns across different languages.  It is also important to highlight that the human JND happens to produce a very good agreement with experimental data, and \textit{this is not a trivial finding, since different JNDs can perform much worse}. Said in other words, the model is in fact sensitive to the shape of the JND, and the human JND happens to produce a small discrepancy between the simulations and the WCS data, as shown in Fig.~\ref{f:nwcs_jnds}.
\subsection*{Departures from universality}

%\begin{figure}[!h]
%\begin{center}
%\includegraphics*[width=0.55\textwidth]{centroid_histograms.eps}
%\end{center}
%\caption{{\bf Numerical World Color Survey with human and neutral
%    JNDs: structure of categories.} The two solid curves represent the
%  histograms of position of category-centroids obtained in simulations with
%  the human (black curve) and neutral JND (blue curve). As a
%  reference, the inverse of the human JND, rescaled by a constant
%  factor, is displayed (red dashed curve). In the central region
%  strong correlation is seen between the centroid distribution from
%  ``human-like'' simulations and the inverse of the human JND, while
%  the outcome of unbiased simulations is flat in the same
%  region. Strong oscillations near the two extrema ($0$ and $1$) are
%  appreciated in both models, typical of ``hard boundaries''. The
%  solid circles displayed at the bottom of the graph represent the
%  ``average pattern'' of the $\sim 150$ populations which display $14$
%  categories at the end of ``human-like'' simulations.}
% \label{f:structure}
% \end{figure}

In the previous section we focused on the question: how sensitive is  the result of our model to the particular form of JND curve? It  turned out that the form of the JND function matters, i.e., choices  different from the ``human'' one do not agree, in statistical terms  (clustering of languages), with the empirical WCS data. However we cannot underestimate the {\em dynamical} and {\em stochastic} nature  of our model: indeed, it does not simply optimise some sort of energy landscape associated to the external constraint (e.g. the JND curve). On the contrary, it goes through a sequence of trials, errors, and successes that are not easily forgotten and affect the subsequent evolution of the category system, as pointed out by a recent study based on time-dependent statistical correlation  functions \cite{mukherjee2011aging}. Such a history-dependent  ``cultural'' process leads to a final state not necessarily optimal  with respect to the JND function: for this reason we consider the JND to be  a {\em weak bias} on category formation, as it is not sufficient  to determine the category pattern. Let us now address this problem in a quantitative way, focusing on the structure of categories, as represented by their centroids.  

\begin{figure}
\begin{center} \includegraphics*[width=0.55\textwidth]{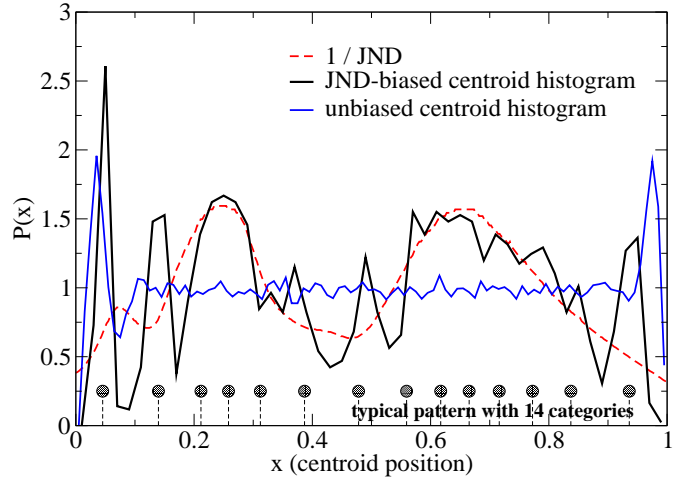} \end{center} 
\caption{{\bf Numerical World Color Survey with human and neutral
    JNDs: structure of categories.} The two solid curves represent the
  histograms of position of category-centroids obtained in simulations with
  the human (black curve) and neutral JND (blue curve). As a
  reference, the inverse of the human JND, rescaled by a constant
  factor, is displayed (red dashed curve). In the central region
  strong correlation is seen between the centroid distribution from
  ``human-like'' simulations and the inverse of the human JND, while
  the outcome of unbiased simulations is flat in the same
  region. Strong oscillations near the two extrema ($0$ and $1$) are
  appreciated in both models, typical of ``hard boundaries''. The
  solid circles displayed at the bottom of the graph represent the
  ``average pattern'' of the $\sim 150$ populations which display $14$
  categories at the end of ``human-like'' simulations.}
 \label{f:structure}
 \end{figure}

Our first aim is to verify a correlation between the average pattern of categories and the JND curve. We simulated $600$ independent populations each with $N=200$ agents. All the simulations were stopped at $10^6$ games per agent, when a quasi-stationary state with high communicative success has been reached. For each population $j\in[1,600]$, we identified the $m_j$ categories shared by the $N$ agents: their centroids are averaged across the agents, giving rise to the population category pattern $s_j=\{x^j_1,x^j_2,...,x^j_{m_j}\}$ where $x^j_i\in(0,1)$ is the average centroid of the $i$-th category of the $j$-th population. Then we collected together in a single set $s$ all the positions $x_i^j$ (with $i\in[1,m_j]$ and $j\in[1,600]$) of centroids from all categories of all populations. Finally we divided the perceptual space, i.e. the segment $[0,1]$ into $50$ bins and count how many centroid's positions fall in each bin, producing the category histogram. This histogram, normalised in order to give total area $1$, is shown in the main graph of Fig.~\ref{f:structure}. We consider such a histogram, also called {\em distribution} in the following, as a good estimate of the probability of finding a category centroid in the Category Game informed with a given JND curve. The black curve is the distribution of centroid positions in populations simulated with the human JND, the blue one is obtained with a neutral (flat) JND. The red dashed curve is the inverse of the human JND curve, rescaled by a constant factor (for the purpose of displaying it on the same scale). We observe that - in the central region $\sim [0.2,0.8]$ of the light spectra - a correlation exists between the probability of finding a centroid and the inverse of the JND curve. Centroids are more easily found where the JND is smaller, i.e., where the perceptual resolution power of agents is higher. Interestingly, large oscillations appear near the two extremes ($0$ and $1$) in both human and neutral simulations.  Such oscillations signal the presence of hard boundaries: indeed no centroids can appear too close to $1$ or $0$ (precisely at  distance smaller than the JND in the extrema, which is $\sim 0.02$  for humans and $\sim 0.01$ in the neutral model) and therefore an almost periodic pattern is likely to appear near the boundaries. This effect resembles the oscillations of density for liquids of hard spheres near a wall \cite{hansen2006liquids}. It is expected that simulations with periodic boundary conditions will not display those oscillations.

%\begin{figure}[!t]
%\begin{center}
%\includegraphics*[width=0.55\textwidth]{distances_new2.eps}
%\end{center}
%\caption{{\bf Numerical World Color Survey with human JNDs: weakness
%    of the bias.} The histogram of the squared distance
%  $d=\sum_{i=1}^{14} (x_i-\overline{x_i})^2$ between the position of
%  the $i$-th centroid and its average (``typical'') value. Black data
%  represent the statistics of those populations displaying $14$
%  linguistic categories at the end of Category Game simulations with
%  the human JND, which were roughly $\sim 1300$ over the total $5000$
%  considered populations. The inset displays the actual values of $d$
%  for each population. Green and red data come from a ``Random'' model
%  where each ``language'' is produced by a uniform random distribution
%  of $14$ category centroids. A random case with a very large number
%  of languages (red curve) represents the ``ideal'' statistics of such
%  a model. The distances $d$ from the average pattern (shown in the inset)
%  appear to have a larger average $\overline{d}_{rand}$ than the
%  Category Game model $\overline{d}_{CG}$ (where by $\overline{d}$ we 
%  refer to the mean value of the random and the CG cases, respectively), but since we are interested
%  in the fluctuations we have rescaled the random model data, dividing
%  them by $\overline{d}_{rand}/\overline{d}_{CG}$, in order to compare
%  the histograms. A power law fit $\sim x^{-3}$ is also shown as a
%  guide for the eye.}
% \label{f:weakness}
% \end{figure}
 
Our next step is estimating how large the influence is of the JND on the structure of categories. For this purpose, we restricted our focus to simulations with the human JND. We sampled $5000$ different populations. Populations with the human JND tend to have a number of categories between $9$ and $19$: this number shows a roughly bell-shaped distribution, whose maximum ($\sim 1300$ populations out of the $5000$ simulated) is at $m=14$.  We restricted further our analysis considering only those populations with $m=14$ categories, in order to single out the largest available group of populations ending up with a given $m$. We computed the average, or ``typical'', pattern of categories for this sub-group, $\overline{s}=\{\overline{x}_1,\overline{x}_2,...,\overline{x}_{14} \}$. The $14$ average positions of typical centroids are displayed at the bottom of Fig.~\ref{f:structure}. 

\begin{figure}
\begin{center} \includegraphics*[width=0.55\textwidth]{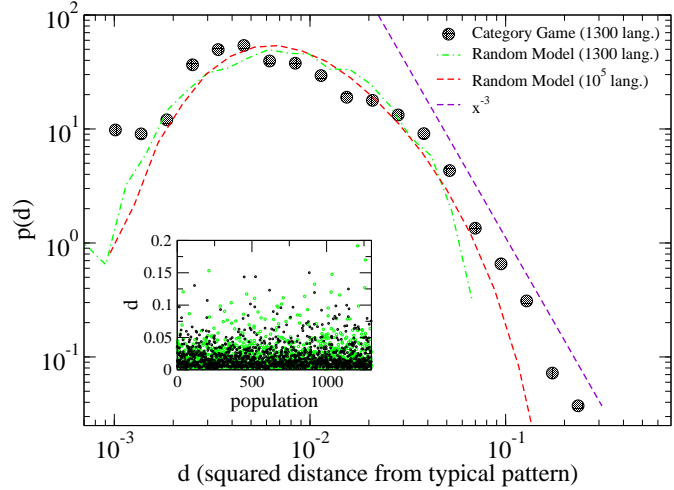} \end{center} 
\caption{{\bf Numerical World Color Survey with human JNDs: weakness
    of the bias.} The histogram of the squared distance
  $d=\sum_{i=1}^{14} (x_i-\overline{x_i})^2$ between the position of
  the $i$-th centroid and its average (``typical'') value. Black data
  represent the statistics of those populations displaying $14$
  linguistic categories at the end of Category Game simulations with
  the human JND, which were roughly $\sim 1300$ of the total $5000$
  considered populations. The inset displays the actual values of $d$
  for each population. Green and red data come from a ``Random'' model
  where each ``language'' is produced by a uniform random distribution
  of $14$ category centroids. A random case with a very large number
  of languages (red curve) represents the ``ideal'' statistics of such
  a model. The distances $d$ from the average pattern (shown in the inset)
  appear to have a larger average $\overline{d}_{rand}$ than the
  Category Game model $\overline{d}_{CG}$ (where by $\overline{d}$ we 
  refer to the mean value of the random and the CG cases, respectively), but since we are interested
  in the fluctuations we have rescaled the random model data, dividing
  them by $\overline{d}_{rand}/\overline{d}_{CG}$, in order to compare
  the histograms. A power law fit $\sim x^{-3}$ is also shown as a
  guide for the eye.}
 \label{f:weakness}
 \end{figure}

Having established the {\em typical pattern}, we went further, asking how far the pattern $s_j$ of the $j$-th population is from the typical one $\overline{s}$. The simplest tool to this end is the Euclidean (squared) distance $d_j=|s_j-\overline{s}|^2\equiv\sum_{i=1}^{14}(x_i^{(j)}-\overline{x}_i)^2$ of the $j$-th population from the typical pattern. The $\sim 1300$ values of $d_j$ obtained are shown in the inset graph of Fig.~\ref{f:weakness}. While it is intuitively clear that such a distance can take values distinctly larger than zero, a quantitative assessment of this observation can be obtained by computing the distribution of the values of $d_j$, which we report in the main graph of Fig.~\ref{f:weakness}. 
%It appears that on a decade of large values (in between $\sim 0.01 \pm 0.1$) the distribution is quite broad: a possible power-law decay $P(d) \sim d^{-\alpha}$ (with $\alpha \simeq 3$) could approximate the fat tail. 
The distribution appears quite broad: indeed its tail at large values (i.e. $d>0.01$) decays slower than an
exponential. In particular, a possible power-law decay $P(d) \sim d^{-\alpha}$ (with $\alpha \simeq 3$) could
approximate such a fat tail.

We stress that a power-law tail is a typical signal of fluctuations {\em larger than normal}. In the absence of strong correlations, the random wandering of stochastic variables (such as the distance from the average that we are considering) is usually expected to have Gaussian or exponential tails. A simple example is borrowed from the theory of fluctuations in equilibrium statistical physics: a Brownian particle in a harmonic potential displays fluctuations of its distance $x$ from the well's center distributed as $\sim e^{-c x^2}$ where $c$ is a constant depending on the temperature and the curvature of the potential. Einstein's celebrated fluctuation theory~\cite{einstein1910} generalises this observation and predicts again an exponential distribution $\sim e^S$ with $S$ the system's entropy, which is the quantity maximised at equilibrium (i.e. $-S$ is the optimised cost). Non-exponential distributions typically reveal that fluctuations do not come from a system which tries to optimise a cost function.

Nevertheless, with our available statistics, the ``fat'' (power-law) nature of our tail cannot be certified. More compelling is the comparison between the results from the Category Game and those from a ``Random model'' where the $14$ centroids are independently and uniformly distributed on the segment $[0,1]$. On one hand the red dashed curve in Fig.~\ref{f:weakness}, coming from a simulation of $10^5$ random languages, illustrates the ``ideal'' distribution of such a model, which has an exponential cut-off for large distances. On the other hand the green curve allows us to evaluate also finite size effects, by taking into account only $1300$ random languages: in such a case the histogram is noisy but close to the ideal one and it is not as broad as the one related to the Category Game.
 
In conclusion we find that in the Category Game - with a non-negligible probability - a population may display a category pattern quite far from the typical one. Thus, ``typical'' does not mean ``certain''. The bias induced by the external constraint (the human JND) is not strong enough to attract all patterns to a typical configuration, and the history-dependent dynamics may lead a population into states only weakly affected by the JND. Repeating such an analysis with different choices of $m\neq 14$ leads to similar results with compatible power-law exponents, signalling the robustness of the result. It might be interesting to repeat the study of these fluctuations of ``distances'' from the typical pattern with the actual data of the WCS. However the number of available languages is too small, and restricting the analysis to a given number of colour terms would reduce the statistics even more.  For these reasons we did not pursue this idea.

\section*{Discussion}

We have analysed the mechanism through which a weak cognitive or perceptual bias ends up influencing the structure of a language, focusing on the prototypical case of colour categorisation and studying the Category Game model. Previous studies showed that in this framework the presence of the JND bias triggers the emergence of universal colour naming patterns whose statistical properties are in agreement with the ones observed in the WCS \cite{cg_pnas_2}, and that the model is able to reproduce the observed hierarchy of colour names \cite{cg_pnas_vittorio}. Here we have seen that:

\begin{enumerate}

\item The particular form assumed by the perceptual bias (i.e., the shape of the JND function) does affect the properties of the universality patterns generated by the model. The fact that the human JND is responsible for a quantitative agreement with the WCS data (see discussion above and \cite{cg_pnas_2},\cite{cg_pnas_vittorio}) is therefore genuinely significant.

\item A further source of biases arises from the dynamical evolution of the system. While the JND is universal and would tend to result in highly similar category patterns in different languages, the specific evolution of each language acts in the opposite direction by introducing frozen accidents and randomness in the process. Quantitatively, the distribution of the distances from the typical pattern of the various languages exhibits a fat tail, which indicates the possibility of significant, culture-dependent, deviations. Our study of fluctuations suggests that colour categorisation is not a simple optimisation process where some cost function is minimised.

\end{enumerate}

These conclusions arise from a series of numerical experiments in which the agents were endowed with artificially manipulated JNDs. 
%It is worth stressing that, at least to our knowledge, there are no previous attempts to connect experimental data (i.e., the ones of the WCS) to a single characteristic (i.e., the roughness) of a \textit{specific} human cognitive feature (i.e., the JND) through a quantitative test of a computational model. 
In this respect, a remark is in order on the agreement with the data obtained when the human JND is considered (at least as far as the test introduced in \cite{kay2003rqc} and the hierarchy of colour names are considered). Whether this is an (extremely) fortunate coincidence or has more profound reasons can not be ``proved" in this context. The task of simulations is that of testing theoretical hypotheses, showing whether they are realistic, and checking for the implications \cite{frankfurt}. 

What we have shown is that a perceptual bias \textit{can} in fact induce universality, to be intended in a statistical sense, and that the shape of this bias can influence the degree of universality (i.e., regularity across different languages). The quite large cross-linguistic variability of the colour naming patterns emergent in population endowed with the human JND indicates that strong deviations from the most common patterns are possible. The broadness of the fluctuations distribution suggests that its origin could be dynamical, i.e., related to a single population's history. This clarifies why universality has to be intended in a statistical sense, and that the history of a language can reduce the tendency of a shared bias to enforce uniformity.

% Do NOT remove this, even if you are not including acknowledgments.

\section*{Acknowledgments}

The authors are grateful to Animesh Mukherjee and Francesca Tria for helpful discussions. V.L. acknowledges support from the EveryAware European project nr. 265432 under FP7-ICT-2009-C and DRUST project funded by the European Science Foundation under EuroUnderstanding Collaborative Research Projects and the KREYON project funded by the Templeton Foundation under contract N. 51663. SONY-CSL provided support in the form of salaries for author VL, but did not have any additional role in the study design, data collection and analysis, decision to publish, or preparation of the manuscript. The specific roles of the authors are articulated in the "author contributions" section. The funders had no role in study design, data collection and analysis, decision to publish, or preparation of the manuscript.

\clearpage

\section*{Appendix: The Category Game model}

The game is played by a population of $N$ individuals. 
Each individual is characterised by its partition of the $[0,1)$ perceptual channel in non-overlapping contiguous segments henceforth called \textit{(perceptual) categories}.  Each category has an associated inventory of words, that constitutes its linguistic counterpart. Individuals are endowed with the capability of transmitting words to each other and to interact non-linguistically through pointing at objects in the environment.

At each time step $t=1,2,..$ two individuals are randomly extracted and interact, one playing as speaker and the other one as hearer. They face a  \textit{scene} of $M$ \textit{objects}, i.e. $M$ real numbers randomly extracted from the interval $[0,1)$, with $M \geq 2$. One of the objects is the topic $h$ that the speaker will try to communicate to the hearer. 

The interaction involves the following steps:

\begin{description}
 \item[Discrimination] \ \\
The speaker perceives the scene, i.e. assigns each object $i \in [0,1)$ of the scene to one of its categories. The category associated with object $i$ is the unique segment $[l,r)$ of the individual perceptual channel for which it holds $l \leq i < r$. An object $k$ is said to be \textit{discriminated} by category $C$ if $k$ is the only object of the scene to be associated to $C$. In other words, if $k$ is discriminated by $C$, then given any object $j$ in the scene it holds $ j \in C \Leftrightarrow j \equiv k$ .

There are two possibilities for the topic $h$:  

\begin{description}
 \item[Either] $h$ is already discriminated by a category $C$;
 \item[or] $h$ and a non empty set $O$ of different objects $i$ fall in the same category $C$. 
\end{description}

In the latter case the speaker refines the category partition of its perceptual channel to discriminate the topic. Given the two objects $a,b$ for which it holds $a \equiv \max_{i \in O} \{i : i < h \}$ and $b \equiv \min_{i \in O} \{ i : i > h \}$,  category $C$ is split in new categories by the introduction of new boundaries in $(a+h)/2$ and $(h+b)/2$. [If $h>i, \; \forall i$ ($h<i, \; \forall i$) then $a$ ($b$) is not defined, and of course the corresponding new boundary is not created.] Each new category inherits the linguistic inventory of $C$, plus a brand new word.

 \item[Word transmission] \ \\
	
After discrimination, the speaker transmits a word to the hearer in order to identify the topic. 

\begin{description}
 \item[If] a previous successful communication event has occurred with the discriminating category the speaker transmits the word that yielded that success;
 \item [else] the speaker transmits the brand new word added to the discriminating category when it was created.
\end{description}

\item[Word reception] \ \\

The hearer receives the transmitted word, and, looking at its repertoire, identifies the set of all categories

\begin{description}
 \item[(i)] whose inventories contain the transmitted word and
 \item[(ii)] that are associated to at least one object in the scene.	
\end{description}

\item[Guessing and outcomes of the game] \ \\

There are now several mutually exclusive possibilities for the hearer:

\begin{description}
 \item[a] The set is empty;
 \item[b] The set contains only one category, corresponding to a single object in the scene;
 \item[c] The set contains only one category, corresponding to more than one object in the scene;
 \item[d] The set contains more than one category.	
\end{description}

Then:

\begin{description}
 \item[if a] The hearer cannot infer which is the topic, and communicates its perplexity to the speaker (we can image that individuals have a built in conventionalised way of doing that, for instance pointing at the sky).
 \item[if b] There is only a candidate object for the hearer, who points at it; 
 \item[if c or d] The hearer points randomly at one of its candidate objects. 
\end{description}

At this point the speaker unveils the topic (pointing at it), and both individuals become aware of the result of their interaction, that is

\begin{description}
 \item[success] if the object pointed by the hearer corresponds to the topic or
 \item[failure] in all the other cases.
\end{description}

% Moreover:

% \begin{description}
%  \item[if the topic was not discriminated by the hearer] The hearer discriminates the topic following the discrimination procedure described above.
% \end{description}

\item[Updating] \ \\

Independently of the outcome of the game, the hearer checks whether the topic is discriminated by one of its categories. If this is not the case, it discriminates the topic following the discrimination procedure described above.

Then:

\begin{description}
 \item[in case of failure] the hearer adds the transmitted word to the category discriminating the topic;
 \item[in case of success] both agents delete all the words but the transmitted one from the inventory of the category discriminating the topic.
\end{description}

\end{description}

\clearpage 

%\section*{References}

% Either type in your references using
% \begin{thebibliography}{}
% \bibitem{}
% Text
% \end{thebibliography}
%
% OR
%
% Compile your BiBTeX database using our plos2009.bst
% style file and paste the contents of your .bbl file
% here.
%\bibliographystyle{plos2015}
%\bibliography{plos_bib}

% This section is for figure legends only, do not include
% graphics in your manuscript file.
%
%\begin{figure}
%\caption{
%{\bf Bold the first sentence.}  Rest of figure caption.  
%}
%\label{Figure_label}
%\end{figure}

%\section*{Tables}
% 
% See introductory notes if you wish to include sideways tables.
%
% NOTE: Please look over our table guidelines at http://www.plosone.org/static/figureGuidelines#tables to make sure that your tables meet our requirements. Certain types of spacing, cell merging, and other formatting tricks may have unintended results and will be returned for revision.
%
%\begin{table}[!ht]
%\caption{
%\bf{Table title}}
%\begin{tabular}{|c|c|c|}
%table information
%\end{tabular}
%\begin{flushleft}Table caption
%\end{flushleft}
%\label{tab:label}
% \end{table}

%\section*{Supporting Information Legends}
%
% Please enter your Supporting Information captions below in the following format:
%\item{\bf Figure SX. Enter mandatory title here.} Enter optional descriptive information here.
% 
%\begin{description}
%\item {\bf}
%\item {\bf}
%\end{description}

\end{document}